\documentclass[journal]{IEEEtran}

\usepackage[
  colorlinks=true,
  linkcolor=blue,
  citecolor=blue,
  urlcolor=blue
]{hyperref}

\usepackage{booktabs}
\usepackage{pgfplots}
\usepackage[linesnumbered,ruled,lined]{algorithm2e}
\usepackage{enumitem}
\usetikzlibrary{shapes.multipart,intersections}
\usepackage{cite}
\usepackage{amsmath,amssymb,amsfonts,amsthm,steinmetz}
\usepackage{mathrsfs}  
\usepackage{textcomp}
\usepackage{acronym}
\usepackage{xcolor}
\usepackage{upgreek,xspace}
\usepackage{array}
\usepackage{tikz}
\usetikzlibrary{calc}
\makeatletter
\newcommand{\gettikzxy}[3]{%
  \tikz@scan@one@point\pgfutil@firstofone#1\relax
  \edef#2{\the\pgf@x}%
  \edef#3{\the\pgf@y}%
}
\makeatother

\usepackage[draft]{todonotes}   
\usepackage[T1]{fontenc}

\usepackage{esvect}

\usepackage{times}
\usepackage{bm}
\usepackage{stmaryrd}
\usepackage{babel}
\usepackage{graphics, graphicx}
\usepackage{gensymb}
\usepackage{cite}
\usepackage{enumitem}
\usepackage{url}

\usepackage[all=normal,paragraphs=normal,floats=tight,mathspacing=normal,wordspacing=normal,charwidths=tight,mathdisplays=normal,leading=normal]{savetrees}

\begin{document}

\title{RIS-Aided Wireless Multiport Sensing: \\Multiplexed De-Embedding of an \\RIS-Programmable Over-the-Air Fixture}

\author{Philipp~del~Hougne,~\IEEEmembership{Member,~IEEE}
\thanks{This work was supported in part by the ANR France 2030 program (project ANR-22-PEFT-0005), the ANR PRCI program (project ANR-22-CE93-0010), the European Union's European Regional Development Fund, and the French Region of Brittany and Rennes M\'etropole through the contrats de plan \'Etat-R\'egion program (projects ``SOPHIE/STIC \& Ondes'' and ``CyMoCoD'').}
\thanks{
P.~del~Hougne is with Univ Rennes, CNRS, IETR - UMR 6164, F-35000, Rennes, France (e-mail: philipp.del-hougne@univ-rennes.fr).
}
}

\maketitle

\begin{abstract}
Wireless multiport sensing aims to remotely retrieve the scattering matrix of a multiport device under test (DUT) connected to not-directly-accessible (NDA) antennas, based on scattering parameters measured with remotely located accessible antennas that couple over the air (OTA) to the NDA antennas. A central bottleneck is that the required number of accessible antennas grows with the number of unknowns in the DUT's scattering matrix. Here, we address this bottleneck by using a reconfigurable intelligent surface (RIS) to generate measurement diversity during the DUT characterization. We interpret the setup as measuring the DUT via an RIS-programmable OTA fixture. We first characterize the RIS-programmable OTA fixture using a specific known tunable load network to terminate the NDA antennas. Then, we reconstruct the DUT's scattering matrix by multiplexed de-embedding of the RIS-programmable OTA fixture, using measurements of the DUT acquired across multiple RIS configurations. Based on our system model formulated in terms of multiport-network theory, we quantify and maximize the diversity of our measurement sequence by optimizing the deployed ensemble of RIS configurations. We validate our approach experimentally at 2.45~GHz in a rich-scattering radio environment. We systematically examine the influence of the number of accessible antennas, the number of RIS configurations, and the optimization of the RIS configurations. For a SISO link (two accessible antennas), optimized RIS configurations reduce the median reconstruction mean-squared error for 100 DUT measurements by 45\% from $1.57{\times}10^{-3}$ to $8.59{\times}10^{-4}$. These results demonstrate that RIS diversity can compensate for a limited number of accessible antennas, paving the way toward low-cost wireless multiport sensing for industrial, biomedical, and smart-environment RFID systems.
\end{abstract}


\section{Introduction}
\label{sec_introduction}

The modulation of an antenna's termination with a tunable load underpins a wide range of techniques for backscatter communications and sensing. Exploited at least since the 1940s~\cite{stockman1948communication,brooker2013lev}, backscatter techniques are nowadays widespread via RFID technology. Moreover, modern programmable-metasurface concepts like reconfigurable intelligent surfaces (RISs) can be understood as arrays of individually tunable backscatter elements. 

Within the realm of wireless backscatter sensing, one can broadly distinguish between digital and analog backscatter modulation~\cite{Marrocco}. Analog backscatter modulation can be further classified according to whether the structure or the load of the backscatter antenna is modulated. In the former case, one seeks to retrieve the reflection coefficient of the backscatter antenna whereas in the latter case one seeks to retrieve the reflection coefficient of the load. Both cases can be generalized to multi-port backscatter modulation, where the ports of an array of backscatter elements are terminated by a tunable load network (which can in principle be fully connected, i.e., it is not necessarily limited to being an ensemble of individual tunable loads). Multi-port backscatter schemes are sometimes referred to as RFID grids~\cite{Marrocco_RFID_GRID,Marrocco_RFID_GRID2}. Ensembles of closely placed individual backscatter elements constitute multi-port backscatter systems because their electromagnetic interactions prevent treating each of them individually without accounting for the others~\cite{marrocco2008multiport,mughal2023statistical,nanni2024stackedRFID,barbot2025differential}. Moreover, some backscatter systems are conceived explicitly as multi-port systems~\cite{caizzone2011multi}. In multi-port backscatter wireless sensing, one seeks to retrieve either the scattering matrix of the antenna array or of the load network. In the special case in which the loads are known to be individual, the load network has a diagonal scattering matrix. 

The problem of remotely estimating the reflection coefficient of a backscatter antenna, relevant when the structure of the backscatter antenna is modulated for analog wireless sensing, has received considerable attention in the literature~\cite{garbacz1964determination,mayhan1994technique,pursula2008backscattering,Capstick2009,bories2010small,van2020verification,sahin2021noncontact,kruglov2023contactless,barbotAutoTune}.
Many of these approaches idealize the wireless propagation environment (WPE), most commonly by assuming free-space propagation. The multi-port generalization of this problem, i.e., remotely estimating the scattering matrix of an array of backscatter antennas, has also been studied~\cite{wiesbeck1998wide,monsalve2013multiport,denicke2012application,del2024virtual,shilinkov2024antenna,del2024virtual2p0,tapie2025scalable,del2025virtual3p0,del2025virtual3p1}; Sec.~II of~\cite{del2025virtual3p0} provides a detailed comparison of the corresponding techniques. The broadest formulation is provided by the ``Virtual Vector Network Analyzer'' (Virtual VNA) framework~\cite{del2024virtual,del2024virtual2p0,tapie2025scalable,del2025virtual3p0,del2025virtual3p1} which supports both closed-form and gradient-based estimation in general WPEs, permits the known terminations to differ from calibration-standard loads, can exploit an arbitrary number of probing antennas, and, in its gradient-based form, is compatible with non-coherent detection.

The problem of remotely estimating the reflection coefficient of the load of a backscatter antenna, relevant when the load of the backscatter antenna is modulated for analog wireless sensing, has also received some attention~\cite{chen2012wireless,bjorninen2011wireless,akbar2015rfid,skrobacz2024new,chen2012coupling,vena2024backscatter,del2025wireless,barbot2025general}. Most existing contributions address restricted versions of this inverse problem by prescribing the propagation conditions or requiring prior knowledge of the backscatter antenna, for example its characteristics or perfect matching. Two independent studies in 2025 removed these restrictions~\cite{del2025wireless,barbot2025general}. While~\cite{barbot2025general} formulates the single-port problem for a backscatter-modulated SISO link using Green's equation,~\cite{del2025wireless} adopts a general multiport-network formulation; the former is equivalent to the single-port SISO specialization of the latter. The genuinely multi-port case had previously been considered in the context of RFID grids~\cite{Marrocco_RFID_GRID,Marrocco_RFID_GRID2}, but these approaches recovered only a vector proportional to the magnitudes of the diagonal entries of the load network's admittance matrix. Consequently, the unknown proportionality factors and all off-diagonal terms remained unresolved. In contrast,~\cite{del2025wireless} enabled the full scattering matrix of a multiport load network to be retrieved unambiguously without requiring special knowledge of the antennas or the WPE.

The approach taken by~\cite{del2025wireless} frames wireless multiport sensing as measuring the device under test (DUT) via an over-the-air (OTA) fixture that needs to be de-embedded. The OTA fixture is thus first characterized with a known tunable load network (TLN) using a Virtual-VNA-type procedure and is subsequently de-embedded from measurements acquired with the unknown DUT~\cite{del2025wireless}. A limitation of this approach is that one fixture realization must expose sufficiently many independent observables to identify $\mathbf{S}^{\mathrm{DUT}}$; the required number of accessible antennas can therefore grow rapidly with the number of DUT ports. This bottleneck was removed in~\cite{del2026low} by retaining the TLN between the OTA fixture and the DUT and reconfiguring it across measurements. The resulting ``multiplexed de-embedding'' jointly recovers the DUT from measurements obtained through multiple known programmable-fixture realizations, none of which needs to be individually informative enough~\cite{del2026low}. This principle is conceptually related to computational imaging with configurational diversity that distributes complementary information across successive measurements, but it accommodates a generally coupled DUT, a nonlinear multiple-scattering forward model, and no sparsity assumption. Experimentally, this low-complexity approach enabled the ten independent complex-valued entries of a reciprocal four-port DUT to be identified from SISO transmission measurements~\cite{del2026low}.

In this paper, we investigate a complementary means of generating the configurational diversity required to overcome the same bottleneck imposed by a limited number of accessible antennas. We use the TLN only to characterize the OTA fixture; during DUT characterization, we reconfigure an RIS embedded in the propagation environment to generate an ensemble of RIS-programmed OTA-fixture realizations. In this way, we gain access to a large configuration space from which we can randomly or judiciously select RIS configurations to improve the conditioning of the DUT identification. This approach is particularly attractive when programmable scattering elements are already available in a smart radio environment. We use the term RIS in a broad sense: we do not require the tunable elements to be colocated on a single surface; in principle, they can be distributed throughout the environment and, for example, implemented based on individually programmable RFID tags~\cite{Bletsas_RFID_RIS,lestini2025modeling}. 

Our general approach is to first characterize the RIS-programmable OTA fixture using a known TLN before characterizing the DUT via multiplexed de-embedding based on measurement diversity provided by the RIS. Prior work has treated the two ingredients of the first step separately: wireless multiport sensing has so far used known TLN realizations to characterize a static OTA fixture with only a limited, operationally irrelevant ambiguity~\cite{del2025wireless,del2026low}, whereas proxy multiport-network parameter estimation for RIS-parametrized radio environments accommodates unknown RIS load characteristics which lead to substantially more multiport-network parameter ambiguities~\cite{sol2024experimentally,del2025experimental,del2025segmented,del2025reducedrank,del2026cross}. Here, we combine both settings since our present problem involves simultaneously known TLN realizations and known RIS control vectors (but unknown RIS load characteristics). We discuss the resulting ambiguity structure in detail in Sec.~\ref{subsec_OTA_Charac_ProblemStatement}.

Our contributions are summarized as follows:
\begin{enumerate}
    \item We introduce RIS-aided wireless multiport sensing, in which a DUT's scattering matrix is estimated via multiplexed de-embedding based on measurements of the DUT via a sequence of realizations of a previously characterized RIS-programmable OTA fixture.
    \item Based on multiport-network theory (MNT), we formulate the characterization of an OTA fixture that is jointly parametrized by known RIS control vectors (but unknown RIS load characteristics) and known TLN realizations. Thereby we combine two previously separate settings of proxy MNT model parameter estimation.
    \item We formulate RIS-based multiplexed DUT de-embedding after OTA-fixture characterization, where the measurement diversity is supplied by RIS configurations (rather than TLN configurations as in~\cite{del2026low}).
    \item We derive a measure of the measurement diversity based on the effective rank of the relevant Jacobian, and we use it to optimize ensembles of RIS configurations for DUT characterization.
    \item We experimentally validate our approach at 2.45~GHz in a rich-scattering environment using reciprocal four-port DUTs and a RIS comprising 15 1-bit-programmable elements. We systematically quantify how the reconstruction accuracy depends on the accessible-antenna count, the number of RIS configurations, and the optimization of the utilized ensemble of RIS configurations.
\end{enumerate}

The remainder of this paper is organized as follows. In Sec.~\ref{sec_SystemModel}, we describe our multiport-network system model. In Sec.~\ref{sec_ProblemStatement}, we formulate the problem statements for the characterization of the RIS-programmable OTA fixture and the subsequent multiplexed de-embedding of the DUT. In Sec.~\ref{sec_Method}, we present our methods for solving these two inverse problems and for optimizing ensembles of RIS configurations. In Sec.~\ref{sec_ExpVal}, we describe our experimental setup and procedure, evaluate the characterization of the RIS-programmable OTA fixture, analyze the resulting measurement diversity, and assess the multiplexed de-embedding of the DUT. Finally, we conclude in Sec.~\ref{sec_conclusion}.

\section{System Model}
\label{sec_SystemModel}

\begin{figure}
\centering
    \includegraphics[width=0.9\columnwidth]{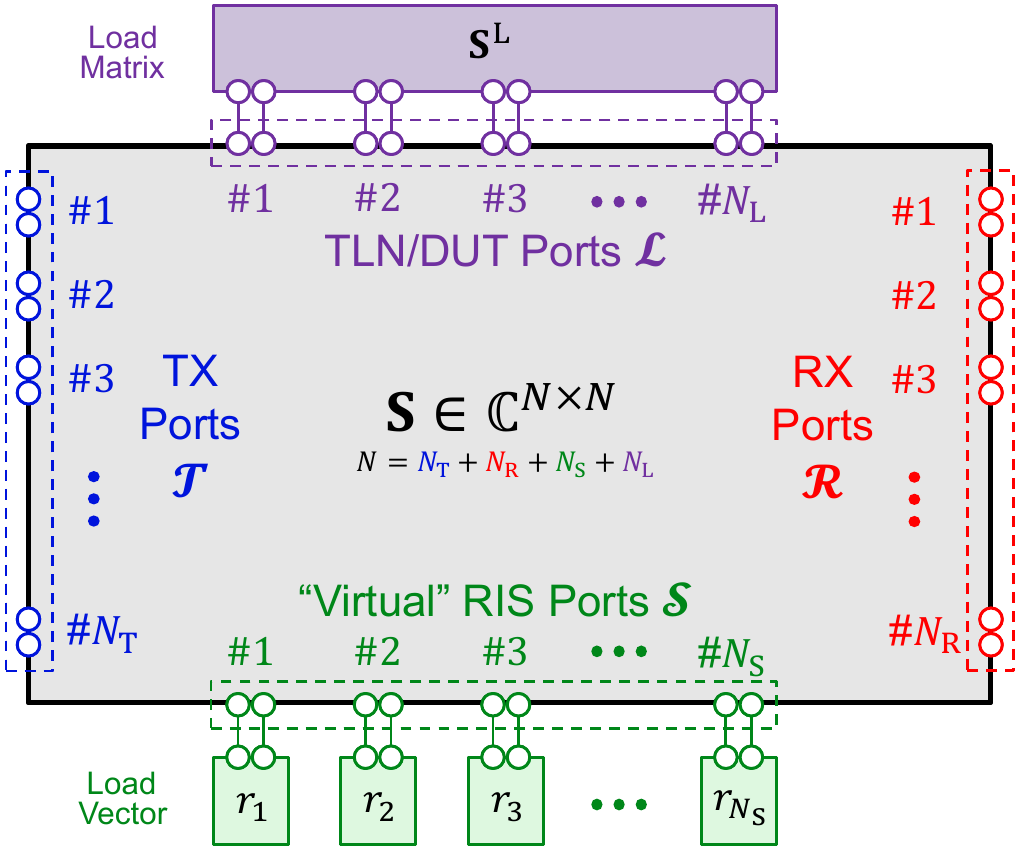}
    \caption{MNT system model.}
    \label{Fig_MNT}
\end{figure}

The considered system is a radio environment comprising $N_\mathrm{T}$ transmitting antennas, $N_\mathrm{R}$ separate receiving antennas, $N_\mathrm{S}$ RIS elements, and $N_\mathrm{L}$ not-directly-accessible (NDA) backscatter antennas terminated by a known TLN realization or the unknown DUT. A schematic of our system model based on MNT is shown in Fig.~\ref{Fig_MNT}. Each RIS element is modeled as an antenna element whose ``virtual'' port is terminated by an individually tunable load. We partition the system into a static subsystem and a tunable subsystem.
The static subsystem is characterized by its scattering matrix
$\mathbf{S}\in\mathbb{C}^{N\times N}$, where
$N=N_\mathrm{T}+N_\mathrm{R}+N_\mathrm{S}+N_\mathrm{L}$.
$\mathbf{S}$ describes scattering by all static components of the system, including antenna structural scattering and scattering by objects in the environment.
The tunable subsystem comprises the $N_\mathrm{S}$ individual tunable loads as well as the $N_\mathrm{L}$-port load network terminating the NDA ports. 
The ensemble of individual tunable loads is characterized by a scattering matrix $\mathbf{\Phi}=\mathrm{diag}(\mathbf{r})\in\mathbb{C}^{N_\mathrm{S}\times N_\mathrm{S}}$, where $\mathbf{r}=[r_1, r_2, \dots, r_{N_\mathrm{S}}]^\top \in\mathbb{C}^{N_\mathrm{S}}$ is the RIS load vector with $r_i$ being the reflection coefficient of the load associated with the $i$th RIS element. 
The load network terminating the NDA ports is characterized by a scattering matrix $\mathbf{S}^\mathrm{L}\in\mathbb{C}^{N_\mathrm{L}\times N_\mathrm{L}}$, which corresponds either to a known TLN realization or to the unknown DUT. 
Thus, the tunable subsystem is characterized by a scattering matrix $\mathbf{\Psi}=\mathrm{blkdiag}(\mathbf{\Phi},\mathbf{S}^\mathrm{L})\in\mathbb{C}^{(N_\mathrm{S}+N_\mathrm{L})\times(N_\mathrm{S}+N_\mathrm{L})}$. We define all scattering matrices and reflection coefficients using a port reference impedance of $50\ \Omega$.

We denote the sets of port indices associated with transmitting antennas, receiving antennas, RIS elements, and NDA antennas by $\mathcal{T}$, $\mathcal{R}$, $\mathcal{S}$, and $\mathcal{L}$, respectively.
Assuming signal generators and signal detectors matched to the $50\ \Omega$ reference impedance, standard MNT~\cite{anderson_cascade_1966,ha1981solid,prod2024efficient} yields the end-to-end channel matrix
$\mathbf{H}\in\mathbb{C}^{N_\mathrm{R}\times N_\mathrm{T}}$
from the transmitting antennas to the receiving antennas as
\begin{equation}
    \mathbf{H}
    = \mathbf{S}_\mathcal{RT}
    + \mathbf{S}_\mathcal{RQ}
      \left( \mathbf{I}_{N_\mathrm{Q}} - \mathbf{\Psi}\,\mathbf{S}_\mathcal{QQ} \right)^{-1} \mathbf{\Psi}\,
      \mathbf{S}_\mathcal{QT},
    \label{eqMNT}
\end{equation}
where $\mathcal{Q}=\mathcal{S}\cup\mathcal{L}$ and $N_\mathrm{Q}=N_\mathrm{S}+N_\mathrm{L}$.
The ports indexed by $\mathcal{Q}$ are ordered as
$(\mathcal{S},\mathcal{L})$, consistently with the block ordering in
$\mathbf{\Psi}$. For any two port-index sets $\mathcal{X}$ and $\mathcal{Y}$, $\mathbf{S}_{\mathcal{XY}}$ denotes the submatrix of $\mathbf{S}$ whose rows and columns are indexed by $\mathcal{X}$ and $\mathcal{Y}$, respectively. Moreover, $\mathbf{I}_a$ denotes the $a\times a$ identity matrix.

Our RIS uses PIN diodes, which are 1-bit-programmable tunable lumped elements, implying $r_i\in\{\alpha,\beta\} \,\forall\,i$, where $\alpha\in\mathbb{C}$ and  $\beta\in\mathbb{C}$ are the reflection coefficients of the two possible load states. Assuming identical PIN diodes for all elements, the RIS load vector is determined by a binary RIS control vector $\mathbf v\in \{0,1\}^{N_\mathrm S}$ via an affine encoding:
\begin{equation}
\mathbf r(\mathbf v)
=
\alpha\mathbf 1_{N_\mathrm S}
+
(\beta-\alpha)\mathbf v,
\label{eq_RIS_encoding}
\end{equation}
where $\mathbf{1}_{N_\mathrm S}$ denotes the $N_\mathrm{S}$-element all-ones vector.
Thus, we have $\mathbf{\Phi}(\mathbf{v})=\mathrm{diag}(\mathbf{r}(\mathbf{v}))$.
 
Given our ultimate goal of estimating $\mathbf{S}^\mathrm{DUT}$, it is helpful to view the system as an RIS-programmable OTA fixture via which we measure the DUT. To make this interpretation more explicit, we first terminate and eliminate the RIS ``virtual'' ports using $\mathbf\Phi(\mathbf v)$. Let
$\mathcal{P}=\mathcal{T}\cup\mathcal{R}\cup\mathcal{L}$ denote the set of remaining ports, with
$N_\mathrm{P}=N_\mathrm{T}+N_\mathrm{R}+N_\mathrm{L}$. 
The reduced scattering matrix
$\overline{\mathbf{S}}(\mathbf v)\in\mathbb{C}^{N_\mathrm{P}\times N_\mathrm{P}}$
of the RIS-programmed OTA fixture associated with the RIS control vector
$\mathbf v$ is obtained by standard MNT port reduction~\cite{anderson_cascade_1966,ha1981solid,prod2024efficient} as
\begin{equation}
    \overline{\mathbf{S}}(\mathbf v)
    =
    \mathbf{S}_{\mathcal{PP}}
    +
    \mathbf{S}_{\mathcal{PS}}
    \left(
        \mathbf{I}_{N_\mathrm{S}}
        -
        \mathbf{\Phi}(\mathbf v)\mathbf{S}_{\mathcal{SS}}
    \right)^{-1}
    \mathbf{\Phi}(\mathbf v)
    \mathbf{S}_{\mathcal{SP}}.
    \label{eq_RIS_reduced_S}
\end{equation}
Upon terminating the NDA ports by $\mathbf{S}^\mathrm{L}$, standard MNT~\cite{anderson_cascade_1966,ha1981solid,prod2024efficient} yields the end-to-end channel matrix in the form
\begin{equation}
    \mathbf{H}\!\left(\mathbf v,\mathbf{S}^\mathrm{L}\right)
    =
    \overline{\mathbf{S}}_{\mathcal{RT}}(\mathbf v)
    +
    \overline{\mathbf{S}}_{\mathcal{RL}}(\mathbf v)
    \left(
        \mathbf{I}_{N_\mathrm{L}}
        -
        \mathbf{S}^\mathrm{L}
        \overline{\mathbf{S}}_{\mathcal{LL}}(\mathbf v)
    \right)^{-1}
    \mathbf{S}^\mathrm{L}
    \overline{\mathbf{S}}_{\mathcal{LT}}(\mathbf v).
    \label{eq_RIS_parametrized_OTA_fixture}
\end{equation}
Whenever the required inverses exist,
\eqref{eq_RIS_parametrized_OTA_fixture}
is algebraically equivalent to \eqref{eqMNT}.
The former is obtained by first eliminating the RIS ports and subsequently eliminating the NDA ports, whereas the latter eliminates both sets of terminated ports jointly.
The interpretation of the system as an RIS-programmable OTA fixture through which the DUT is measured is made explicit in~\eqref{eq_RIS_parametrized_OTA_fixture}.

\section{Problem Statement}
\label{sec_ProblemStatement}

Our ultimate goal is to estimate
$\mathbf{S}^\mathrm{DUT}$
based on $K_\mathrm{DUT}$ measurements of
$\mathbf{H}(\mathbf{v}^{(k)},\mathbf{S}^\mathrm{DUT})$
for a set of known RIS control vectors
$\{\mathbf{v}^{(k)}\}_{k=1}^{K_\mathrm{DUT}}$.
This estimation can be understood as a ``multiplexed de-embedding'' of the RIS-programmable OTA fixture, analogous to the ``multiplexed de-embedding'' of a TLN-programmable OTA fixture in~\cite{del2026low}.
Here, the attribute ``multiplexed'' emphasizes that information about the same unknown DUT is distributed across measurements acquired through multiple RIS-programmed realizations
$\{\overline{\mathbf{S}}(\mathbf{v}^{(k)})\}_{k=1}^{K_\mathrm{DUT}}$
of the OTA fixture and that all of these measurements are jointly exploited to estimate
$\mathbf{S}^\mathrm{DUT}$.
Consequently, an individual RIS configuration need not provide sufficiently many independent measurements to identify the DUT on its own; instead, the required information can be supplied collectively by the RIS-induced fixture diversity.

A prerequisite for this multiplexed de-embedding is a sufficiently unambiguous characterization of the RIS-programmable OTA fixture. By ``sufficiently unambiguous,'' we mean that the calibrated fixture model can accurately predict
$\mathbf H(\mathbf v,\mathbf S^\mathrm L)$
for every admissible pair
$(\mathbf v,\mathbf S^\mathrm L)$.
Residual parameter ambiguities that do not affect the mapping
$(\mathbf v,\mathbf S^\mathrm L)\mapsto
\mathbf H(\mathbf v,\mathbf S^\mathrm L)$
are operationally irrelevant.
Thus, the RIS-programmable OTA fixture must be calibrated before the DUT's scattering characteristics can be estimated by multiplexed de-embedding. During this calibration stage, $K_\mathrm{cal}$ measurements of
$   \mathbf{H}\!\left(
        \mathbf{v}^{(c)},
        \mathbf{S}^{\mathrm{TLN},(c)}
    \right)$ 
are acquired for known RIS control vectors
$\mathbf{v}^{(c)}$
and known TLN scattering matrices
$\{\mathbf S^{\mathrm{TLN},(c)}\}_{c=1}^{K_\mathrm{cal}}$.

Overall, we thus successively tackle two inverse problems.
\textit{First}, we characterize the RIS-programmable OTA fixture using measurements obtained with known TLN realizations and known RIS control vectors.
\textit{Second}, after replacing the known calibration TLN by the unknown DUT, we estimate $\mathbf{S}^\mathrm{DUT}$ using measurements obtained with the unknown DUT and known RIS control vectors.
We formalize these two problems in the following subsections.

\subsection{Characterization of the RIS-Programmable OTA Fixture}
\label{subsec_OTA_Charac_ProblemStatement}

To start, we position the considered characterization problem relative to three related lines of prior work.
\textit{First}, the ``Virtual VNA'' technique aims to determine the scattering matrix of a DUT by terminating a subset of NDA DUT ports with known loads, using at least three distinct and known terminations per NDA port~\cite{del2024virtual,del2024virtual2p0,del2025virtual3p0,del2025virtual3p1,tapie2025scalable}. Using only individual load terminations generally leaves ambiguities that can be removed with suitably chosen coupled loads~\cite{del2024virtual2p0,del2025virtual3p0,del2025virtual3p1}.
\textit{Second}, prior work on wireless multi-port sensing applies the ``Virtual VNA'' principle to characterize a static OTA fixture using known TLN realizations~\cite{del2025wireless,del2026low}. Since a coupled load connecting an accessible and an NDA port is impractical in the scenario of wireless multi-port sensing, one operationally irrelevant ambiguity remains~\cite{del2025wireless,del2026low}.
\textit{Third}, proxy MNT parameter estimation for RIS-parametrized radio environments addresses the case in which the RIS control vector is known but the RIS load characteristics are unknown~\cite{sol2024experimentally,del2025experimental,del2025segmented,del2025reducedrank,del2026cross}. The load characteristics are assumed unknown because in practice the RIS design may be proprietary or the available component specifications may be insufficient to determine the load reflection coefficients. Moreover, the RIS elements' loads are often only 1-bit-programmable and there are typically no coupled loads.\footnote{Even ``beyond-diagonal'' RISs typically rely on individual tunable lumped elements such that they admit a physics-consistent description based on a \textit{diagonal} tunable load matrix~\cite{del2025physics}.} 
Consequently, the resulting proxy MNT parameters exhibit substantial ambiguities
(see the Appendix in~\cite{del2026mimobounds} for an overview of the ambiguity classes);
nonetheless, these ambiguities are operationally irrelevant when the estimated proxy MNT parameter set accurately reproduces the physical system's observable mapping from every admissible RIS control vector to the corresponding end-to-end channel matrix.
Analogous proxy-MNT parameter estimation techniques have also been developed for dynamic metasurface antennas~\cite{tapie_DMAMNTcal,tapie2026channel}.

Against this background, the present characterization problem combines these prior settings: one subset of ports is terminated by unknown 1-bit-programmable individual loads, whereas another subset is terminated by a known TLN.
More precisely, we know the RIS control vectors but not the complex reflection coefficients associated with the two RIS states, while we know the scattering matrices of the TLN realizations.
The known TLN realizations therefore impose additional constraints on the proxy-model parameters associated with the NDA ports.
We seek proxy MNT parameters that satisfy these constraints and accurately reproduce the complete observable mapping
$(\mathbf v,\mathbf S^\mathrm L)\mapsto\mathbf H(\mathbf v,\mathbf S^\mathrm L)$
over the considered class of admissible RIS control vectors and terminations of the NDA ports.
We do \textit{not} seek an unambiguous reconstruction of all internal MNT model parameters.

We parameterize the proxy MNT model by
$\boldsymbol{\theta}_\mathrm{OTA}
\triangleq
(\mathbf H_0,\mathbf A,\mathbf\Gamma,\mathbf B,\tilde{\alpha},\tilde{\beta})$,
where
$\mathbf H_0\in\mathbb C^{N_\mathrm R\times N_\mathrm T}$,
$\mathbf A\in\mathbb C^{N_\mathrm R\times N_\mathrm Q}$,
$\mathbf\Gamma\in\mathbb C^{N_\mathrm Q\times N_\mathrm Q}$,
$\mathbf B\in\mathbb C^{N_\mathrm Q\times N_\mathrm T}$,
and
$\tilde{\alpha},\tilde{\beta}\in\mathbb C$.
Within this proxy model,
$\mathbf H_0$,
$\mathbf A$,
$\mathbf\Gamma$,
$\mathbf B$,
$\tilde{\alpha}$, and
$\tilde{\beta}$
play the roles of
$\mathbf S_{\mathcal{RT}}$,
$\mathbf S_{\mathcal{RQ}}$,
$\mathbf S_{\mathcal{QQ}}$,
$\mathbf S_{\mathcal{QT}}$,
$\alpha$, and
$\beta$, respectively.
We do not require the proxy parameters to coincide with a uniquely identifiable physical parameterization of the RIS-programmable OTA fixture.
We exploit the known reciprocity of the RIS-parametrized OTA fixture by imposing $\mathbf\Gamma=\mathbf\Gamma^\top$.

For a candidate parameter tuple
$\boldsymbol{\theta}_\mathrm{OTA}$,
we define the proxy RIS load matrix as
\begin{equation}
    \widetilde{\mathbf\Phi}(\mathbf v)
    \triangleq
    \operatorname{diag}\!\left[
        \tilde{\alpha}\mathbf 1_{N_\mathrm S}
        +
        (\tilde{\beta}-\tilde{\alpha})\mathbf v
    \right],
    \label{eq_OTA_proxy_Phi}
\end{equation}
and the scattering matrix of the tunable subsystem as
\begin{equation}
    \widetilde{\mathbf\Psi}(\mathbf v,\mathbf S^\mathrm L)
    \triangleq
    \operatorname{blkdiag}\!\left[
        \widetilde{\mathbf\Phi}(\mathbf v),
        \mathbf S^\mathrm L
    \right].
    \label{eq_OTA_proxy_Psi}
\end{equation}
The proxy MNT model then predicts
\begin{equation}
    \mathbf H\!\left(
        \mathbf v,
        \mathbf S^\mathrm L;
        \boldsymbol{\theta}_\mathrm{OTA}
    \right)
    =
    \mathbf H_0
    +
    \mathbf A
    \left[
        \mathbf I_{N_\mathrm Q}
        -
        \widetilde{\mathbf\Psi}(\mathbf v,\mathbf S^\mathrm L)
        \mathbf\Gamma
    \right]^{-1}
    \widetilde{\mathbf\Psi}(\mathbf v,\mathbf S^\mathrm L)
    \mathbf B.
    \label{eq_OTA_proxy_forward_model}
\end{equation}

Let
$\mathbf H_\mathrm{phys}(\mathbf v,\mathbf S^\mathrm L)$
denote the end-to-end channel matrix produced by the physical fixture.
In the ideal noise-free setting, we call
$\boldsymbol{\theta}_\mathrm{OTA}$
operationally equivalent to the physical fixture if
\begin{equation}
    \mathbf H\!\left(
        \mathbf v,
        \mathbf S^\mathrm L;
        \boldsymbol{\theta}_\mathrm{OTA}
    \right)
    =
    \mathbf H_\mathrm{phys}\!\left(
        \mathbf v,
        \mathbf S^\mathrm L
    \right),
    \
    \forall\,(\mathbf v,\mathbf S^\mathrm L)
    \in\mathcal D_\mathrm{OTA},
    \label{eq_OTA_operational_equivalence}
\end{equation}
where $\mathcal D_\mathrm{OTA}$ denotes the considered class of admissible RIS control vectors and load-network terminations of the NDA ports.
In practice, finite and noisy measurements only allow us to approximate and empirically assess the condition in~\eqref{eq_OTA_operational_equivalence}.

Let
$\mathbf H_\mathrm{cal}^{(c)}$
denote the end-to-end channel matrix measured from the physical fixture for the known calibration pair
$(\mathbf v^{(c)},\mathbf S^{\mathrm{TLN},(c)})$,
for $c=1,\ldots,K_\mathrm{cal}$.
We estimate the proxy MNT parameters $\widehat{\boldsymbol{\theta}}_\mathrm{OTA}$ by solving
\begin{equation}
    \widehat{\boldsymbol{\theta}}_\mathrm{OTA}
    \in
    \underset{
        \boldsymbol{\theta}_\mathrm{OTA}
        \in\mathcal A_\mathrm{OTA}
    }{\operatorname{arg\,min}}
    \;
    \frac{
        \displaystyle
        \sum_{c=1}^{K_\mathrm{cal}}
        \left\|
            \mathbf H_\mathrm{cal}^{(c)}
            -
            \mathbf H\!\left(
                \mathbf v^{(c)},
                \mathbf S^{\mathrm{TLN},(c)};
                \boldsymbol{\theta}_\mathrm{OTA}
            \right)
        \right\|_\mathrm F^2
    }{
        \displaystyle
        \sum_{c=1}^{K_\mathrm{cal}}
        \left\|
            \mathbf H_\mathrm{cal}^{(c)}
        \right\|_\mathrm F^2
    },
    \label{eq_OTA_characterization_problem}
\end{equation}
where $\mathcal A_\mathrm{OTA}$ denotes the set of parameter tuples satisfying
$\mathbf\Gamma=\mathbf\Gamma^\top$,
$\tilde{\alpha}\neq\tilde{\beta}$,
and the requirement that all inverses in~\eqref{eq_OTA_proxy_forward_model}
exist for the considered RIS control vectors and load-network terminations.
We may additionally impose
$\tilde{\alpha}=0$
as a gauge-fixing convention, consistently with prior proxy-MNT estimation approaches~\cite{del2025experimental,del2025segmented,del2025reducedrank,del2026cross,tapie_DMAMNTcal}.

The solution of~\eqref{eq_OTA_characterization_problem} need not be unique.
We regard this nonuniqueness as operationally irrelevant when the resulting parameter tuples induce the same observable mapping, as formalized by~\eqref{eq_OTA_operational_equivalence}.
We therefore assess the characterization of our RIS-parametrized OTA fixture using
$K_\mathrm{test}$
held-out unseen test measurements
$(\mathbf v_\mathrm{test}^{(t)},\mathbf S_\mathrm{test}^{\mathrm{TLN},(t)})$,
$t=1,\ldots,K_\mathrm{test}$,
which we do not use to solve~\eqref{eq_OTA_characterization_problem}.
We quantify the calibrated proxy model's prediction accuracy on the unseen test examples with
\begin{equation}
    \zeta_\mathrm{OTA}
    \triangleq
    \frac{
        \left\langle
        \operatorname{SD}_{t}
        \!\left(
            \left[\mathbf H_\mathrm{test}^{(t)}\right]_{ij}
        \right)
        \right\rangle_{i,j}
    }{
        \left\langle
        \operatorname{SD}_{t}
        \!\left(
            \left[
                \mathbf H_\mathrm{test}^{(t)}
                -
                \mathbf H\!\left(
                    \mathbf v_\mathrm{test}^{(t)},
                    \mathbf S_\mathrm{test}^{\mathrm{TLN},(t)};
                    \widehat{\boldsymbol{\theta}}_\mathrm{OTA}
                \right)
            \right]_{ij}
        \right)
        \right\rangle_{i,j}
    },
    \label{eq_OTA_zeta}
\end{equation}
where $\operatorname{SD}_{t}(\cdot)$ is evaluated across the held-out test measurements for each end-to-end channel coefficient $(i,j)$, and $\langle\cdot\rangle_{i,j}$ denotes averaging over the considered end-to-end channel coefficients.
This definition compares the typical held-out channel-variation scale with the typical residual prediction-error scale.
Averaging the realization-wise standard deviations over channel coefficients avoids mixing variation across different channel coefficients with variation across held-out realizations, which would otherwise make SISO and MIMO cases inequivalent under the normalization.
A larger value of $\zeta_\mathrm{OTA}$ indicates a more accurate proxy MNT model.

\subsection{DUT De-Embedding through the RIS-Programmable OTA Fixture}
\label{subsec_MultiplexedDeembedding_ProblemStatement}

After characterizing the RIS-programmable OTA fixture, we replace the known calibration TLN by the unknown DUT and acquire
$K_\mathrm{DUT}\geq 1$
end-to-end channel matrices using known RIS control vectors.
This de-embedding problem relates to the two preceding wireless multi-port sensing approaches as follows.
In~\cite{del2025wireless}, a characterized static OTA fixture is de-embedded using the end-to-end channel coefficients available for one fixture realization.
In~\cite{del2026low}, a TLN remains connected between the OTA fixture and the DUT and is reconfigured to generate multiple programmable-fixture realizations whose measurements are jointly de-embedded.
In the present setting, by contrast, the TLN is used only during fixture calibration; when the DUT is measured, the TLN plays no role (it is either removed or it is held static and treated as part of the DUT), while the RIS provides reconfigurability to generate measurement diversity.

Let
$\mathbf H_\mathrm{DUT}^{(k)}$
denote the end-to-end channel matrix measured with the unknown DUT connected to the NDA ports and with the known RIS control vector
$\mathbf v^{(k)}$,
for
$k=1,\ldots,K_\mathrm{DUT}$.
Using the previously estimated fixture parameters
$\widehat{\boldsymbol{\theta}}_\mathrm{OTA}$,
we estimate the DUT scattering matrix by solving
\begin{equation}
    \widehat{\mathbf S}^{\mathrm{DUT}}
    \in
    \underset{
        \mathbf Z\in\mathcal A_\mathrm{DUT}
    }{\operatorname{arg\,min}}
    \;
    \frac{
        \displaystyle
        \sum_{k=1}^{K_\mathrm{DUT}}
        \left\|
            \mathbf H_\mathrm{DUT}^{(k)}
            -
            \mathbf H\!\left(
                \mathbf v^{(k)},
                \mathbf Z;
                \widehat{\boldsymbol{\theta}}_\mathrm{OTA}
            \right)
        \right\|_\mathrm F^2
    }{
        \displaystyle
        \sum_{k=1}^{K_\mathrm{DUT}}
        \left\|
            \mathbf H_\mathrm{DUT}^{(k)}
        \right\|_\mathrm F^2
    },
    \label{eq_DUT_deembedding_problem}
\end{equation}
where, under the assumed DUT reciprocity,
$\mathcal A_\mathrm{DUT}$ contains complex-symmetric matrices for which all inverses required by~\eqref{eq_OTA_proxy_forward_model} exist for the employed RIS control vectors.
We use the directly measured ground-truth DUT scattering matrix only to assess the reconstruction accuracy.

A reciprocal
$N_\mathrm L$-port DUT has
$N_\mathrm L(N_\mathrm L+1)/2$
independent complex-valued scattering parameters.
Consequently, when all
$N_\mathrm R N_\mathrm T$
entries of each end-to-end channel matrix are measured, a necessary counting condition for identifying the DUT is
\begin{equation}
    K_\mathrm{DUT}N_\mathrm R N_\mathrm T
    \geq
    \frac{N_\mathrm L(N_\mathrm L+1)}{2}.
    \label{eq_DUT_counting_condition}
\end{equation}
This condition is not necessarily sufficient because the corresponding nonlinear sensitivities may be redundant or poorly conditioned.

Our DUT de-embedding formulation applies to both
$K_\mathrm{DUT}=1$
and
$K_\mathrm{DUT}>1$.
The former corresponds to single-shot de-embedding (similar to~\cite{del2025wireless}); the counting condition in~\eqref{eq_DUT_counting_condition} is a necessary but not sufficient condition for local identifiability in single-shot de-embedding.
For
$K_\mathrm{DUT}>1$,
measurements acquired under multiple RIS configurations are jointly fitted in~\eqref{eq_DUT_deembedding_problem} (similar to~\cite{del2026low}), allowing RIS-induced fixture diversity to compensate for insufficient or redundant sensitivities in individual configurations.
We emphasize that our DUT de-embedding formulation makes no assumptions about whether the RIS configuration or configuration ensemble is selected randomly or via an optimization protocol.

\section{Method}
\label{sec_Method}

We now describe how we successively solve the two inverse problems introduced in Sec.~\ref{sec_ProblemStatement}.
We then describe how we optimize ensembles of RIS configurations for the second inverse problem (i.e., DUT de-embedding).

\subsection{Characterization of the RIS-Programmable OTA Fixture}
\label{subsec_OTA_Charac_Method}

We solve the proxy-MNT characterization problem in~\eqref{eq_OTA_characterization_problem} by jointly optimizing all unknown parameters of
$\boldsymbol{\theta}_\mathrm{OTA} = (\mathbf H_0,\mathbf A,\mathbf\Gamma,\mathbf B,\tilde{\alpha},\tilde{\beta})$ using gradient-based optimization.
The proxy-MNT estimation procedures in~\cite{del2025experimental,del2025segmented,del2025reducedrank,del2026cross,tapie_DMAMNTcal} imposed $\tilde{\alpha}=0$, such that a single reference measurement directly determined $\mathbf H_0$, while $N_\mathrm{S}$ single-toggle measurements determined the directions of the columns and rows of $\mathbf A$ and $\mathbf B$, respectively, up to unknown complex scaling factors.
These closed-form steps are not applicable in the present scenario because their outcomes would depend on the TLN realization.
Consequently, we jointly fit the measured dependence on the RIS configuration and the TLN realization.
We therefore do not impose the gauge $\tilde{\alpha}=0$, although that would remain possible in the present scenario.

We represent every complex-valued optimization variable through its real and imaginary parts.
We enforce reciprocity by parameterizing
$\mathbf\Gamma$
as a complex-symmetric matrix through its independent upper-triangular entries.
We initialize
$\mathbf H_0$
as the sample mean of the measured training end-to-end channel matrices.
We draw the real and imaginary parts of the entries of
$\mathbf A$
and
$\mathbf B$
independently from zero-mean Gaussian distributions with standard deviation $0.03$.
Likewise, we draw the independent upper-triangular entries of
$\mathbf\Gamma$
with standard deviation $0.02$ and construct its lower-triangular part by symmetry.
Moreover, we initialize
$\tilde{\alpha}=0.2$
and
$\tilde{\beta}=-0.2$.

Because~\eqref{eq_OTA_characterization_problem} is nonconvex, we expect some sensitivity to the initialization that we accommodate by performing multiple restarts and retaining the solution that attained the lowest training loss. Since we observe that the initialization sensitivity is much stronger for $N_\mathrm{T}=N_\mathrm{R}=1$ than for $N_\mathrm{T}=N_\mathrm{R}>1$, we use ten restarts for $N_\mathrm{T}=N_\mathrm{R}=1$ and three restarts for $N_\mathrm{T}=N_\mathrm{R}>1$.
For each restart, we minimize the normalized squared-error objective in~\eqref{eq_OTA_characterization_problem} using the Adam optimizer and automatic differentiation. We use full-batch Adam with learning rate
$10^{-3}$
for at most
$25\,000$
optimization steps.
During each restart, we retain the parameter tuple attaining the lowest finite training loss over all iterations rather than necessarily using the final iterate.

To assess the learned proxy model on the held-out test data, we use 
$\zeta_\mathrm{OTA}$
defined in~\eqref{eq_OTA_zeta}.

\subsection{DUT De-Embedding through the RIS-Programmable OTA Fixture}
\label{subsec_MultiplexedDeembedding_Method}

After estimating
$\widehat{\boldsymbol{\theta}}_\mathrm{OTA}$,
we keep all proxy-MNT parameters fixed and solve the DUT de-embedding problem in~\eqref{eq_DUT_deembedding_problem} only with respect to
$\mathbf S^\mathrm{DUT}$.
We enforce DUT reciprocity by parameterizing
$\mathbf S^\mathrm{DUT}$
directly as a complex-symmetric matrix using its independent upper-triangular entries.
We do not otherwise constrain its structure; in particular, we even estimate DUTs expected to be diagonal as general reciprocal multi-port networks.

We represent the real and imaginary parts of the independent DUT parameters as real optimization variables and minimize the normalized channel-prediction error in~\eqref{eq_DUT_deembedding_problem} using full-batch Adam with a learning rate of $10^{-3}$. Because the DUT de-embedding problem is nonconvex, we use five deterministic random initializations and $4000$ optimization steps per initialization. For each initialization, we draw a random complex matrix with scale $0.05$ and symmetrize it. During each initialization, we retain the iterate attaining the lowest finite objective value and ultimately select the result with the lowest objective value across the five initializations.

\subsection{Model-Based Optimization of RIS Configuration Ensembles}
\label{subsec_RIS_configuration_optimization}

The de-embedding problem in~\eqref{eq_DUT_deembedding_problem} does not prescribe how the $K_\mathrm{DUT}$ RIS configurations are selected. Random configurations provide a model-free baseline. However, once we have characterized the RIS-programmable OTA fixture, we can use the calibrated proxy model to select an optimized ensemble of RIS configurations.
While our ultimate objective is to minimize the reconstruction error for the unknown physical DUT, this error cannot be evaluated at the moment of optimizing the RIS configurations because the DUT's scattering matrix is not known. A possible surrogate objective could be the reconstruction error for a representative ensemble of synthetic DUTs. However, evaluating this surrogate objective would require repeatedly solving the nonlinear DUT de-embedding problem during optimization. We therefore use a computationally less expensive surrogate objective that aims to provide more diverse local sensitivities to the independent parameters of the DUT.

The formulation of our surrogate optimization objective builds directly on the Jacobian-based diversity analysis introduced for low-complexity wireless multiport sensing in~\cite{del2026low}. Following~\cite{Blumensath2013},~\cite{del2026low} interpreted the Jacobian of the nonlinear DUT-to-measurement map as a local sensing matrix and used its singular-value spectrum to assess the diversity supplied by randomly chosen programmable-fixture realizations. This Jacobian-based perspective subsequently also inspired the definition of the effective electromagnetic degrees of freedom of backscatter MIMO systems~\cite{del2025effective}. In contrast to the present work,~\cite{del2026low} generated measurement diversity by reconfiguring the TLN rather than an RIS and presented only a preliminary optimization of the effective rank of the Jacobian, without examining whether the optimized configurations improved the DUT reconstruction. Here, we instead optimize RIS configuration ensembles and evaluate the resulting reconstruction accuracy. We optimize a surrogate objective based on the effective rank of the Jacobian, which promotes nonredundant local sensitivity directions. We apply this objective to two candidate domains: a measured pool (MP), which contains a limited set of randomly chosen RIS configurations for which we experimentally measured the DUT response, and the full universe (FU), which contains all $2^{N_\mathrm S}$ admissible binary RIS configurations whose responses are predicted using the calibrated proxy model.

Let
$\mathbf s\in\mathbb C^d$
collect the upper-triangular entries of
$\mathbf S^\mathrm{DUT}$,
where
$d=N_\mathrm{L}(N_\mathrm{L}+1)/2$.
For notational compactness, we introduce
\begin{equation}
    \mathring{\boldsymbol{\Psi}}
    \triangleq
    \widetilde{\mathbf\Psi}
    \left(
        \mathbf v,\mathbf S^\mathrm{DUT}
    \right),
    \ \ \
    \mathbf M
    \triangleq
    \mathbf I_{N_\mathrm Q}
    -
    \mathring{\boldsymbol{\Psi}}\mathbf\Gamma,
    \ \ \
    \mathbf X
    \triangleq
    \mathbf M^{-1}
    \mathring{\boldsymbol{\Psi}}
    \mathbf B,
    \label{eq_DUT_design_auxiliary_quantities}
\end{equation}
such that
$\mathbf H=\mathbf H_0+\mathbf A\mathbf X$.
Using
$\mathrm d\mathbf M^{-1}
=
-\mathbf M^{-1}
(\mathrm d\mathbf M)
\mathbf M^{-1}$~\cite{hjorungnes2007complex},
we differentiate the proxy forward model with respect to
$\mathbf S^\mathrm{DUT}$
(for fixed $\mathbf v$) and obtain
\begin{equation}
    \mathrm d\mathbf H
    =
    \mathbf A\mathbf M^{-1}
    \mathrm d\mathring{\boldsymbol{\Psi}}
    \left(
        \mathbf\Gamma\mathbf X+\mathbf B
    \right).
    \label{eq_DUT_design_differential}
\end{equation}
Since
$\mathring{\boldsymbol{\Psi}}
=
\operatorname{blkdiag}
[
\widetilde{\mathbf\Phi}(\mathbf v),
\mathbf S^\mathrm{DUT}
]$
and the terminated ports are ordered as
$(\mathcal S,\mathcal L)$,
we define
$\mathbf P_\mathrm L
\triangleq
[\mathbf 0_{N_\mathrm L\times N_\mathrm S}\ \mathbf I_{N_\mathrm L}]^\top
\in
\{0,1\}^{N_\mathrm Q\times N_\mathrm L}$,
such that
$\mathrm d\mathring{\boldsymbol{\Psi}}
=
\mathbf P_\mathrm L
\mathrm d\mathbf S^\mathrm{DUT}
\mathbf P_\mathrm L^\top$.
We denote by $\mathbf{0}_{a\times b}$ the $a\times b$ all-zeros matrix.
Upon defining
$\mathbf L
\triangleq
\mathbf A\mathbf M^{-1}\mathbf P_\mathrm L$
and
$\mathbf R
\triangleq
\mathbf P_\mathrm L^\top
(\mathbf\Gamma\mathbf X+\mathbf B)$,
we obtain
\begin{equation}
    \mathrm d\mathbf H
    =
    \mathbf L\,
    \mathrm d\mathbf S^\mathrm{DUT}\,
    \mathbf R,
    \label{eq_DUT_design_reduced_differential}
\end{equation}
which is the counterpart of
[(15),~\cite{del2026low}]
for the calibrated RIS-programmable proxy model.
Unlike [(15),~\cite{del2026low}],
\eqref{eq_DUT_design_reduced_differential}
does not explicitly involve
$(\mathbf S^\mathrm{DUT})^{-1}$
and therefore remains directly evaluable even for a reference DUT with singular scattering matrix.

Let
$\{\mathbf E_\ell\}_{\ell=1}^{d}$
denote the symmetric basis matrices associated with the independent upper-triangular entries of
$\mathbf S^\mathrm{DUT}$.
Specifically, if
$s_\ell$
corresponds to the upper-triangular index
$(i_\ell,j_\ell)$,
then
$\mathbf E_\ell
=
\mathbf e_{i_\ell}\mathbf e_{i_\ell}^\top$
for
$i_\ell=j_\ell$,
whereas
$\mathbf E_\ell
=
\mathbf e_{i_\ell}\mathbf e_{j_\ell}^\top
+
\mathbf e_{j_\ell}\mathbf e_{i_\ell}^\top$
for
$i_\ell<j_\ell$.
We thus have
$\mathbf S^\mathrm{DUT}
=
\sum_{\ell=1}^{d}s_\ell\mathbf E_\ell$
and
$\mathrm d\mathbf S^\mathrm{DUT}
=
\sum_{\ell=1}^{d}\mathbf E_\ell\,\mathrm ds_\ell$.
Substituting the latter expression into~\eqref{eq_DUT_design_reduced_differential} and vectorizing yields
\begin{equation}
  \operatorname{vec}(\mathrm d\mathbf H)
=
\sum_{\ell=1}^{d}
\operatorname{vec}(\mathbf L\mathbf E_\ell\mathbf R)\,\mathrm ds_\ell  .
\end{equation}
Consequently, for a reference DUT with scattering matrix
$\mathbf S_0$,
we define the complex local Jacobian as
\begin{equation}
    \mathbf J
    \left(
        \mathbf v;\mathbf S_0
    \right)
    \triangleq
    \left.
    \frac{
        \partial\operatorname{vec}(\mathbf H)
    }{
        \partial\mathbf s^\top
    }
    \right|_{ \mathbf S_0
    }
    =
    \begin{bmatrix}
        \mathbf j_1 & \cdots & \mathbf j_d
    \end{bmatrix},
    \label{eq_DUT_design_complex_Jacobian}
\end{equation}
where
\begin{equation}
   \mathbf j_\ell
    =
    \left.
    \operatorname{vec}
    \left(
        \mathbf L\mathbf E_\ell\mathbf R
    \right)
    \right|_{
        \mathbf S_0
    }.  
    \label{eq_j_ell}
\end{equation}
The expression for
$\mathbf j_\ell$ in \eqref{eq_j_ell}
is the counterpart of
[(18),~\cite{del2026low}]
for the calibrated RIS-programmable proxy model.

For an ensemble
$\mathcal V_{K_\mathrm{DUT}}
=
(\mathbf v^{(1)},\ldots,\mathbf v^{(K_\mathrm{DUT})})$,
we vertically concatenate the corresponding local Jacobians:
\begin{equation}
    \mathbf J
    \left(
        \mathcal V_{K_\mathrm{DUT}};
        \mathbf S_0
    \right)
    \triangleq
    \begin{bmatrix}
        \mathbf J
        \left(
            \mathbf v^{(1)};
            \mathbf S_0
        \right)^\top
        &
        \cdots
        &
        \mathbf J
        \left(
            \mathbf v^{(K_\mathrm{DUT})};
            \mathbf S_0
        \right)^\top
    \end{bmatrix}^\top.
    \label{eq_DUT_design_stacked_Jacobian}
\end{equation}
Let
$\{\sigma_q\}_{q=1}^{d}$
denote the singular values of
$\mathbf J(\mathcal V_{K_\mathrm{DUT}};\mathbf S_0)$, padded with zeros when the Jacobian has fewer than
$d$ rows.
Analogous to~\cite{del2026low}, we quantify the local measurement diversity through the Jacobian's effective rank~\cite{roy2007effective}:
\begin{equation}
    R_\mathrm{eff}
    \left(
        \mathbf J
    \right)
    \triangleq
    \exp
    \left(
        -\sum_{q=1}^{d}
        p_q\ln (p_q)
    \right),
    \label{eq_DUT_design_effective_rank}
\end{equation}
where $p_q
    \triangleq
    \sigma_q \ \Big/ \left(\sum_{m=1}^{d}\sigma_m\right)$ and we use the convention
$0\ln (0)=0$. For a non-zero Jacobian,
$1\leq R_\mathrm{eff}(\mathbf J)\leq d$.
For a fixed number of nonzero singular values,
$R_\mathrm{eff}(\mathbf J)$
is maximized when they are equal; its global maximum
$d$
is attained when all
$d$
singular values are equal and nonzero.
Because
$R_\mathrm{eff}$
depends on the normalized singular values, it quantifies the balance of the local sensitivity spectrum but not its absolute magnitude.
To efficiently evaluate $R_\mathrm{eff}(\mathbf{J})$, we define the per-configuration Jacobian Gram matrix $    \mathbf G
    \left(
        \mathbf v;
        \mathbf S_0
    \right)
    \triangleq
    \mathbf J
    \left(
        \mathbf v;
        \mathbf S_0
    \right)^\dagger
    \mathbf J
    \left(
        \mathbf v;
        \mathbf S_0
    \right)$,
where $(\cdot)^\dagger$ denotes the conjugate transpose.
Because the ensemble Jacobian is formed by vertical concatenation in~\eqref{eq_DUT_design_stacked_Jacobian},
\begin{equation}
    \mathbf J
    \left(
        \mathcal V_{K_\mathrm{DUT}};
        \mathbf S_0
    \right)^\dagger
    \mathbf J
    \left(
        \mathcal V_{K_\mathrm{DUT}};
        \mathbf S_0
    \right)
    =
    \sum_{k=1}^{K_\mathrm{DUT}}
    \mathbf G
    \left(
        \mathbf v^{(k)};
        \mathbf S_0
    \right).
    \label{eq_DUT_design_ensemble_Gram}
\end{equation}
The singular values required in~\eqref{eq_DUT_design_effective_rank} can be obtained as the square roots of the eigenvalues of this ensemble Jacobian Gram matrix. This reformulation is exact and allows us to precompute the Gram contribution of every candidate configuration for each reference DUT.

We use the median effective rank of the Jacobian over an ensemble of synthetic reference DUTs as a computationally inexpensive surrogate objective. As mentioned earlier, this surrogate objective has the advantages of not requiring knowledge of the unknown DUT's scattering matrix nor requiring repeatedly solving the nonlinear DUT de-embedding problem during optimization.
Related diversity-based surrogate objectives have previously been used in model-agnostic measurement-in-the-loop optimization~\cite{del2019optimally,del2020optimal} and for selecting configurations from finite libraries of premeasured responses~\cite{li2024measurement,zhao2026EuCAP}. The latter approach resembles our MP restriction, but is not model-based because its objective is evaluated from measured response patterns. By contrast, both the MP and FU optimizations considered here evaluate the Jacobian exclusively through the calibrated proxy model. An experimentally calibrated MNT model has recently also enabled model-based diversity optimization for a fabricated dynamic metasurface antenna~\cite{tapie2026optimizing}. As in~\cite{tapie2026optimizing}, we do not claim a formal equivalence between maximizing the surrogate objective and minimizing the final DUT reconstruction error; we assess their relation experimentally below.

To evaluate our surrogate objective, we generate ten fixed synthetic reference DUTs
$\{\mathbf S_r^\mathrm{ref}\}_{r=1}^{10}$.
We generate each reference from an independent random complex-symmetric matrix and rescale it such that its maximum singular value is drawn uniformly from
$[0.35,0.85]$.
This construction enforces reciprocity and strict passivity.
Our surrogate objective is then
\begin{equation}
    \rho
    \left(
        \mathcal V_{K_\mathrm{DUT}}
    \right)
    \triangleq
    \underset{r}{\operatorname{median}}  \
    R_\mathrm{eff}
    \left(
        \mathbf J
        \left(
            \mathcal V_{K_\mathrm{DUT}};
            \mathbf S_r^\mathrm{ref}
        \right)
    \right).
    \label{eq_DUT_design_robust_score}
\end{equation}

We denote the two considered domains for $\mathcal V_{K_\mathrm{DUT}}$ by $\mathcal D_\mathrm{MP}$ and $\mathcal D_\mathrm{FU} = \{0,1\}^{N_\mathrm S}$. The MP domain contains 2500 random RIS configurations for which the DUT responses are experimentally acquired. This restriction allows us to evaluate the DUT reconstruction error for the optimized ensembles without additional experimental measurements, while the selection of the optimized ensembles remains entirely model-based. Meanwhile, the FU domain instead contains all $2^{N_\mathrm S}$ admissible binary configurations. The MP domain is thus a subset of the FU domain: $\mathcal D_\mathrm{MP} \subset \mathcal D_\mathrm{FU}$.

Formally, we solve
\begin{equation}
    \mathcal V_{K_\mathrm{DUT},\mathcal D}^{\star}
     = 
    \underset{
        \mathcal V_{K_\mathrm{DUT}}\in\mathcal A(\mathcal D)
    }{\operatorname{arg\,max}}
    \rho
    \left(
        \mathcal V_{K_\mathrm{DUT}}
    \right),
    \ 
    \mathcal D
    \in
    \{
        \mathcal D_\mathrm{MP},
        \mathcal D_\mathrm{FU}
    \},
    \label{eq_DUT_design_optimization_problem}
\end{equation}
where $\mathcal{A}(\mathcal{D})$ denotes the admissible ensembles of
$K_\mathrm{DUT}$
RIS control vectors drawn from $\mathcal D$.

We solve~\eqref{eq_DUT_design_optimization_problem} independently for each considered value of
$K_\mathrm{DUT}$
using multi-start coordinate-exchange ascent.
During each pass, we visit the
$K_\mathrm{DUT}$
ensemble positions in random order.
At each position, we evaluate all admissible replacements in vectorized batches by subtracting the Gram contribution of the current configuration and adding that of each candidate configuration.
We compute
$\rho$
from the resulting Gram-matrix eigenspectra and accept the replacement yielding the largest increase greater than
$10^{-10}$.
We terminate the search early if a complete pass yields no improvement.

For each value of $K_\mathrm{DUT}$, the MP search starts from six ensembles of cardinality $K_\mathrm{DUT}$: one greedy ensemble, constructed by sequentially adding the RIS configuration that maximizes the current value of $\rho$, and five independently sampled random ensembles. From each initialization, we perform at most four coordinate-exchange passes, where one pass visits each selected RIS configuration and replaces it by the best available candidate if this increases $\rho$. We retain the ensemble with the largest final value of $\rho$. The FU search also starts from six ensembles of cardinality $K_\mathrm{DUT}$: the corresponding MP optimum, and five independently sampled random ensembles. Again, from each initialization, we perform at most four coordinate-exchange passes, retaining the ensemble with the largest final value of $\rho$.

For a given value of $K_\mathrm{DUT}$ and TX-RX subset choice, we benchmark the optimized ensemble of RIS configurations against a random baseline obtained from 500 randomly drawn subsets of the MP domain. For each TX-RX subset choice, we compute $\rho$ for each random ensemble and use the median of these 500 values as the random-baseline value for that TX-RX subset choice.

\section{Experimental Validation}
\label{sec_ExpVal}

In this section, we first describe our experimental setup and procedure in Sec.~\ref{subsec_expSetupProc}. Then, we assess the characterization of the RIS-programmable OTA fixture in Sec.~\ref{subsec_ResultsCharactOTAfixture}, analyze the RIS-induced measurement diversity in Sec.~\ref{subsec_AnalysisPFdiversity}, and evaluate multiplexed DUT de-embedding in Sec.~\ref{subsec_AnalysisDUTdeembedding}.

\subsection{Experimental Setup and Procedure}
\label{subsec_expSetupProc}

\begin{figure}
    \centering
    \includegraphics[width=\columnwidth]{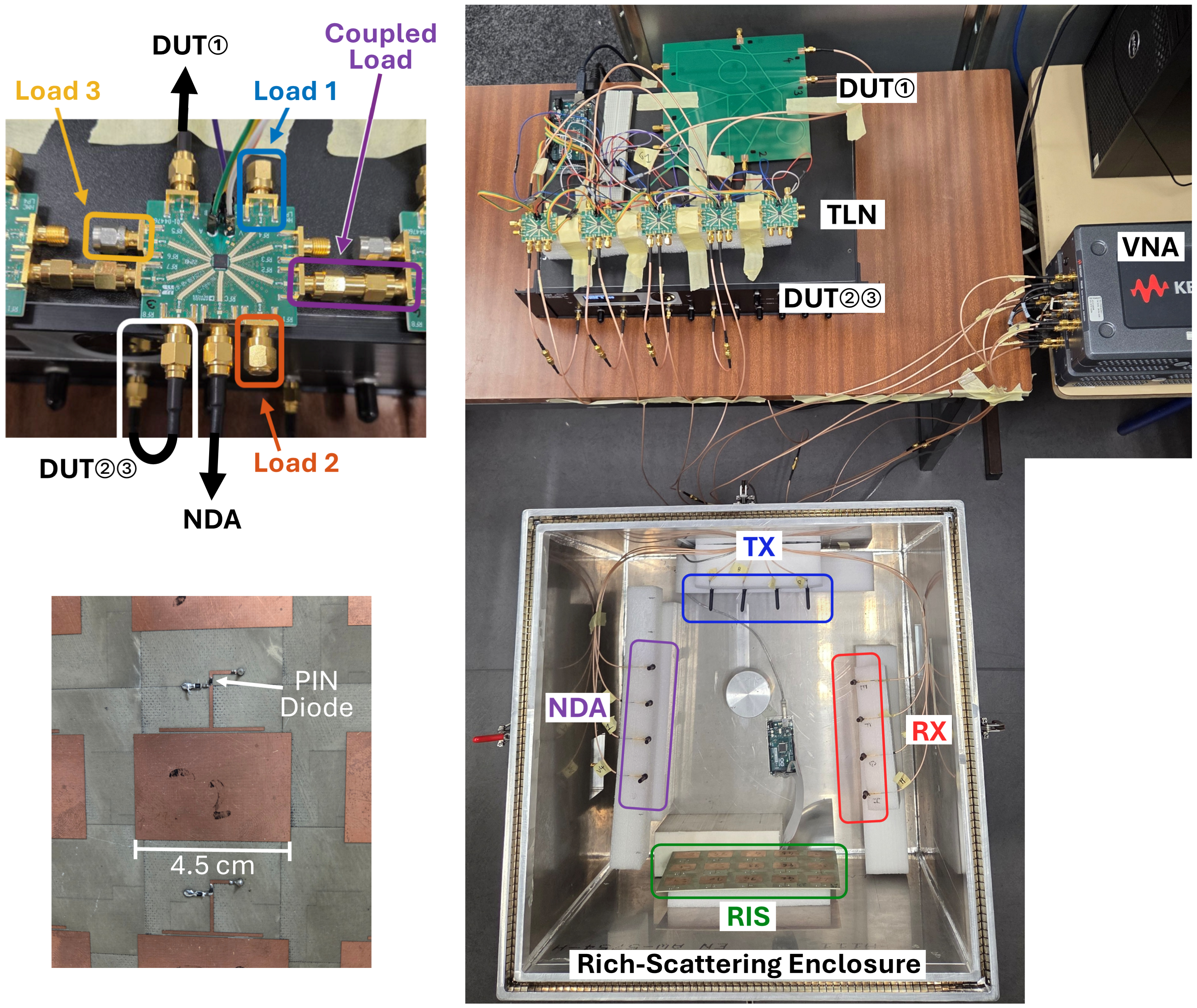}
    \caption{Photograph of the experimental setup (top cover over enclosure removed to show interior). On the left, close-up views of the TLN and the RIS are shown.}    
    \label{Fig2}
\end{figure}

Our experimental setup is displayed in Fig.~\ref{Fig2} and resembles the setups used in~\cite{del2025wireless,del2026low}; the main difference is that our present setup additionally comprises an RIS. Specifically, we consider 4-port, linear, passive, reciprocal, time-invariant DUTs, either a complex transmission-line network or an ensemble of individual delay lines. The DUT is connected via a TLN to four NDA antennas (ANT-24G-HL90-SMA). The four NDA antennas are regularly spaced at roughly half a wavelength (for our operating frequency of 2.45~GHz). The TLN is based on four SP8T switches (HMC321AL94E). Each switch connects an NDA antenna port either to a DUT or a known termination. The known terminations comprise three individual loads (open circuit, short circuit, matched) and coupled loads between neighboring switches. The individual loads are \textit{not} required to be calibration standards, and ours are not once propagation through the switches is included. Throughout this work, we treat the switch and associated coaxial cables as part of the DUT and the TLN. The four NDA antennas couple OTA to a set of accessible antennas (ANT-24G-HL90-SMA), comprising four transmitting and four distinct receiving antennas. The two four-element linear arrays have regular half-wavelength spacing and are oriented orthogonal to each other to avoid a strong line-of-sight path. The OTA coupling between the accessible and NDA antennas is highly complicated due to the rich-scattering enclosure of dimensions $59\ \mathrm{cm} \times 60 \ \mathrm{cm} \times 58 \mathrm{cm}$. Moreover, the scattering environment is parametrized by an RIS comprising 15 1-bit-programmable elements. The RIS element's reconfigurability is based on a PIN diode and follows the design used in~\cite{kaina2014hybridized,ahmed2023over}. Importantly, the PIN diode is electrically very small at the operating frequency, which ensures the validity of our model of the PIN diode as a ``virtual'' lumped port terminated by a tunable load.

We begin with a calibration step in which we use a four-port VNA to determine the scattering characteristics of the TLN realizations (in order to subsequently characterize the RIS-programmable OTA fixture) and the DUT (only used as ground truth to assess our results). To that end, we connect the four switch ports to four VNA ports instead of the four NDA antenna ports. 

Next, we connect the switch ports to the NDA antenna ports and we connect eight VNA ports to the eight accessible antennas. Using an eight-port VNA (two cascaded Keysight P5024B 4-port VNAs), we measure the $4\times 4$ end-to-end channel matrix for a given termination of the NDA antennas (either with a known TLN realization or a DUT) and a given RIS control vector. While VNA-based end-to-end channel measurements are the easiest option in our laboratory, our technique does not inherently require a VNA. Moreover, our technique does not inherently require a MIMO setup. In our post-processing analysis, we can later consider a subset of the measured MIMO matrix while assuming the ports of unused accessible antennas were terminated in matched loads. In particular, this allows us to consider 16 SISO scenarios. SISO channel sounding has particularly low RF-chain requirements and can be realized based on common SDR-based setups used in the realm of RFID~\cite{vena2024backscatter}.

Initially, we characterize the RIS-programmable OTA fixture as described in Sec.~\ref{subsec_OTA_Charac_ProblemStatement} and Sec.~\ref{subsec_OTA_Charac_Method}. We repeatedly cycle through the 109 admissible TLN realizations and pair each occurrence of a TLN realization with a randomly drawn RIS control vector. We acquire $K_\mathrm{cal}=1024$ such pairs to estimate the proxy-MNT parameters and an additional $K_\mathrm{test}=109$ pairs for held-out validation. Because the TLN sequence is cycled periodically, the held-out set contains every admissible TLN realization exactly once; in this setup, the resulting combined TLN--RIS pairs are absent from the calibration set. We use the resulting measurements to solve~\eqref{eq_OTA_characterization_problem} and assess the calibrated model through $\zeta_\mathrm{OTA}$ as defined in~\eqref{eq_OTA_zeta}.

Then, we measure each DUT in turn following the multiplexed de-embedding procedure described in Sec.~\ref{subsec_MultiplexedDeembedding_ProblemStatement} and Sec.~\ref{subsec_MultiplexedDeembedding_Method}. We generate one common pool of random RIS configurations and reuse the same pool for every DUT. During these core DUT measurements, all four switches connect the NDA antenna ports to the current DUT; hence, the measurement diversity is generated by the RIS configurations (rather than by cycling through the calibration TLN realizations as in~\cite{del2026low}). We select subsets of $K_\mathrm{DUT}$ measurements from this pool either randomly or through the model-based MP optimization described in Sec.~\ref{subsec_RIS_configuration_optimization}, and use them for multiplexed de-embedding as formulated in~\eqref{eq_DUT_deembedding_problem}. We use the independently measured ground-truth DUT scattering matrices only to evaluate the reconstruction accuracy.

To quantify experimental stability, we interleave monitor-only repeats of three fixed system states: two TLN realizations and one DUT realization, each paired with a distinct fixed RIS control vector. These measurements are excluded from proxy characterization and DUT de-embedding. Let $\mathbf H_{\mathrm{mon},0}^{(q)}$ and $\mathbf H_{\mathrm{mon},1}^{(q)}$ denote the channel matrices from two nominally identical measurements of monitor state $q$. Analogously to the definition of~$\zeta_\mathrm{OTA}$ in~\eqref{eq_OTA_zeta}, we define
\begin{equation}
    \zeta_\mathrm{stability}
    \triangleq
    \frac{
        \left\langle
        \operatorname{SD}_{q}
        \!\left(
            \left[\mathbf H_{\mathrm{mon},0}^{(q)}\right]_{ij}
        \right)
        \right\rangle_{i,j}
    }{
        \left\langle
        \operatorname{SD}_{q}
        \!\left(
            \left[
                \mathbf H_{\mathrm{mon},1}^{(q)}
                -
                \mathbf H_{\mathrm{mon},0}^{(q)}
            \right]_{ij}
        \right)
        \right\rangle_{i,j}
    }.
    \label{eq_stability_zeta}
\end{equation}
When multiple monitor checkpoints are available, we evaluate~\eqref{eq_stability_zeta} separately for each checkpoint and TX-RX subset; we then report the median and 10th-to-90th percentiles, over TX-RX subset choices, of the worst (lowest) checkpoint value for each subset.

\subsection{Characterization of the RIS-Programmable OTA Fixture}
\label{subsec_ResultsCharactOTAfixture}

\begin{table}[b]
    \centering
    \caption{Median and 10th-to-90th percentiles (across all possible TX-RX subset choices) with $K_\mathrm{cal}=1024$ for $\zeta_\mathrm{stability}$ and $\zeta_\mathrm{OTA}$.}
    \label{Table1}
    \begin{tabular}{c c c}
        \hline
        $N_\mathrm{T}=N_\mathrm{R}$ & $\zeta_\mathrm{stability}$ [dB] & $\zeta_\mathrm{OTA}$ [dB] \\
        \hline
        1 & 52.6 \textcolor{gray}{(46.6--55.7)} & 52.3 \textcolor{gray}{(38.4--53.7)} \\
        2 & 53.0 \textcolor{gray}{(49.7--55.1)} & 48.9 \textcolor{gray}{(47.2--49.8)} \\
        3 & 52.7 \textcolor{gray}{(51.4--54.3)} & 47.6 \textcolor{gray}{(46.7--47.8)} \\
        4 & 52.8 \textcolor{gray}{(52.8--52.8)} & 46.8 \textcolor{gray}{(46.8--46.8)} \\
        \hline
    \end{tabular}
\end{table}

We begin by summarizing the stability of our experiments underlying the characterization of the RIS-programmable OTA fixture with $K_\mathrm{cal}=1024$. We compute $\zeta_\mathrm{stability}$ for each sentinel checkpoint and retain the lowest value for each TX-RX subset. We report the median and 10th-to-90th percentiles of these worst-checkpoint values across all possible TX-RX subset choices in the second column in Table~\ref{Table1}. The median values display no significant dependence on the value of $N_\mathrm{T}=N_\mathrm{R}$, ranging from $52.6$ to $53.0$~dB.

Next, for each possible value of $N_\mathrm{T}=N_\mathrm{R}$ and each TX-RX subset choice, we calibrate the OTA-fixture model with all $K_\mathrm{cal}=1024$ training examples and evaluate its prediction accuracy for the held-out test scenarios. We summarize the median and 10th-to-90th percentiles of these test accuracies in the third column of Table~\ref{Table1}. The median $\zeta_\mathrm{OTA}$ decreases moderately as $N_\mathrm{T}=N_\mathrm{R}$ increases, from $52.3$~dB for SISO to $46.8$~dB for $4\times4$ MIMO. The SISO distribution is strongly asymmetric: its median is $52.3$~dB, but the 10th percentile drops to $38.4$~dB while the 90th percentile remains close at $53.7$~dB. Thus, most SISO links are characterized with near-stability-limited accuracy, but a small subset of links is substantially less accurate. The dependence on the chosen TX-RX subset is strongly reduced for larger antenna subsets because each $\zeta_\mathrm{OTA}$ value is computed from multiple channel coefficients, and different large subsets share many of the same TX and RX ports. Consistent with this interpretation, the $3\times3$ case yields $47.6$~dB with a 10th-to-90th percentile interval of only $46.7$--$47.8$~dB. Altogether, all median values of $\zeta_\mathrm{OTA}$ are very high, exceeding 45~dB.

\begin{figure}
    \centering
    \includegraphics[width=0.9\columnwidth]{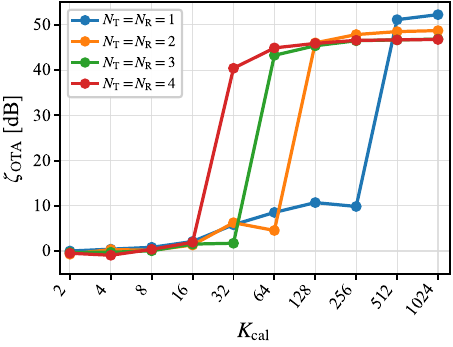}
    \caption{OTA-fixture model accuracy $\zeta_\mathrm{OTA}$ as a function of the number of calibration measurements $K_\mathrm{cal}$, for four different values of $N_\mathrm{T}=N_\mathrm{R}$. For each value of $N_\mathrm{T}=N_\mathrm{R}$, the curve corresponds to the TX-RX subset choice nearest to the median $\zeta_\mathrm{OTA}$ across all possible TX-RX subset choices at $K_\mathrm{cal}=1024$ (which is summarized in Table~\ref{Table1}).}
    \label{Fig3}
\end{figure}

To investigate the minimum required value of $K_\mathrm{cal}$ to achieve a high value of $\zeta_\mathrm{OTA}$, we now perform a $K_\mathrm{cal}$ ablation for one representative TX-RX subset for each value of $N_\mathrm{T}=N_\mathrm{R}$. We choose this representative subset as the one whose $\zeta_\mathrm{OTA}$ at $K_\mathrm{cal}=1024$ is closest to the median $\zeta_\mathrm{OTA}$ across all possible TX-RX subset choices for the same value of $N_\mathrm{T}=N_\mathrm{R}$. The resulting dependence of $\zeta_\mathrm{OTA}$ on $K_\mathrm{cal}$ is shown in Fig.~\ref{Fig3}. We observe a sharp transition between calibration sets that are too small to constrain the inverse problem and calibration sets that yield high prediction accuracy. This transition occurs earlier as $N_\mathrm{T}=N_\mathrm{R}$ increases: for the representative $4\times4$ case, $\zeta_\mathrm{OTA}$ jumps from $1.9$~dB at $K_\mathrm{cal}=16$ to $40.4$~dB at $K_\mathrm{cal}=32$, whereas the representative $3\times3$ and $2\times2$ cases show analogous jumps between $K_\mathrm{cal}=32$ and $64$, and between $K_\mathrm{cal}=64$ and $128$, respectively. The representative SISO case requires substantially more calibration data, with $\zeta_\mathrm{OTA}=9.9$~dB at $K_\mathrm{cal}=256$ and $51.2$~dB at $K_\mathrm{cal}=512$.
This transition scale is also consistent with a simple raw parameter-counting argument. In our setup, the proxy model has $N_\mathrm{Q}=N_\mathrm{S}+N_\mathrm{L}=19$ internal terminated ports. For $N_\mathrm{T}=N_\mathrm{R}=n$, the number of complex proxy-model parameters (before accounting for gauge ambiguities) is $P(n) = n^2 + 2nN_\mathrm{Q} + N_\mathrm{Q}(N_\mathrm{Q}+1)/2 + 2$. Thus, $P(1)=231$, $P(2)=272$, $P(3)=315$, and $P(4)=360$. Since each calibration measurement supplies $n^2$ complex end-to-end channel coefficients, the corresponding raw counting thresholds are $231$, $68$, $35$, and $23$ calibration measurements, respectively. These values are not strict identifiability bounds because the parameterization contains gauge redundancies and the nonlinear sensitivities need not be independent or well conditioned. Nonetheless, we observe that its predicted threshold values align well with the observed transition intervals in Fig.~\ref{Fig3}: $256$--$512$, $64$--$128$, $32$--$64$, and $16$--$32$.  Qualitatively similar threshold-like accuracy jumps were already observed for the RIS-only setting in~\cite{sol2024experimentally}.

We also checked the importance of optimizer restarts for avoiding bad local optima arising from the nonconvex OTA-model calibration problem. For the SISO case, the training loss varies substantially across initializations: the median spread between the best and worst restart losses is $27.9$~dB, and a restart other than the first one is retained in $81\%$ of the cases. For the MIMO cases, selecting the best of three restarts instead of the first restart has negligible effect in almost all cases: the median training-loss reduction is essentially zero, and the 90th percentile remains below $0.01$~dB. The retained best checkpoint usually occurs before the maximum of $25{,}000$ Adam steps; the median iteration indices of the retained checkpoints are $21{,}153$, $13{,}894$, $15{,}322$, and $8{,}742$ for $1\times1$, $2\times2$, $3\times3$, and $4\times4$, respectively. These trends support our choice to use ten restarts for SISO and three restarts for MIMO, as well as the chosen maximum number of $25{,}000$ iterations.

\subsection{Analysis of the RIS-Programmable OTA Fixture's Measurement Diversity}
\label{subsec_AnalysisPFdiversity}

\begin{figure}
    \centering
    \includegraphics[width=\columnwidth]{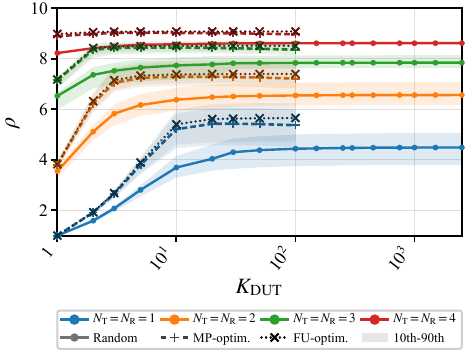}
    \caption{Surrogate objective $\rho$ as a function of the number of DUT measurements $K_\mathrm{DUT}$. Lines show medians, and shaded regions indicate 10th-to-90th-percentile intervals over TX-RX subset choices. Table~\ref{Table2} summarizes the data displayed for $K_\mathrm{DUT}=100$.}
    \label{Fig4}
\end{figure}

\begin{table}[b]
    \centering
    \caption{$\rho$ at $K_\mathrm{DUT}=100$. Entries report medians, and gray intervals indicate 10th-to-90th percentiles over TX-RX subset choices.}    
    \label{Table2}
    \begin{tabular}{c c c c}
        \hline
        $N_\mathrm{T}=N_\mathrm{R}$ & Random & MP-optim. & FU-optim. \\
        \hline
        1 & 4.44 \textcolor{gray}{(3.75--5.01)} & 5.37 \textcolor{gray}{(4.52--6.00)} & 5.66 \textcolor{gray}{(4.72--6.25)} \\
        2 & 6.55 \textcolor{gray}{(6.16--7.06)} & 7.22 \textcolor{gray}{(6.82--7.75)} & 7.40 \textcolor{gray}{(7.05--7.96)} \\
        3 & 7.84 \textcolor{gray}{(7.62--8.14)} & 8.36 \textcolor{gray}{(8.15--8.56)} & 8.52 \textcolor{gray}{(8.31--8.71)} \\
        4 & 8.62 \textcolor{gray}{(8.62--8.62)} & 8.97 \textcolor{gray}{(8.97--8.97)} & 9.08 \textcolor{gray}{(9.08--9.08)} \\
        \hline
    \end{tabular}
\end{table}

As explained in Sec.~\ref{subsec_RIS_configuration_optimization}, we use the metric $\rho$ to quantify the effective dimensionality and balance of the local DUT-sensitivity spectrum supplied by a given ensemble of RIS configurations, and we also use $\rho$ as the surrogate objective for RIS-configuration optimization. The dependence of $\rho$ on $K_\mathrm{DUT}$ is shown in Fig.~\ref{Fig4} for different values of $N_\mathrm{T}=N_\mathrm{R}$ and for random, MP-optimized, and FU-optimized ensembles of RIS configurations. For SISO and $K_\mathrm{DUT}=1$, only a single complex-valued scalar end-to-end channel coefficient is measured, so $\rho$ is unity. Moreover, by definition, $\rho$ is bounded from above by the number of complex DUT degrees of freedom, which is $d=10$ for our reciprocal four-port DUTs.

Increasing $K_\mathrm{DUT}$ generally increases $\rho$, as seen in Fig.~\ref{Fig4}, although strict monotonicity is not guaranteed because the effective rank depends on the full singular-value spectrum of the Jacobian rather than only on its algebraic rank. For random RIS configurations, the curves have essentially saturated by $K_\mathrm{DUT}=100$, and in some cases much earlier. 
At $K_\mathrm{DUT}=100$, the median random values are $4.44$, $6.55$, $7.84$, and $8.62$ for $1\times1$, $2\times2$, $3\times3$, and $4\times4$, respectively, as summarized in Table~\ref{Table2}. 
Thus, increasing $N_\mathrm{T}=N_\mathrm{R}$ increases the effective dimensionality and balance of the local DUT-sensitivity spectrum, as expected, but with diminishing returns as $\rho$ approaches its upper bound. The 10th-to-90th-percentile intervals are widest for SISO, $3.75$--$5.01$, and narrow as $N_\mathrm{T}=N_\mathrm{R}$ increases, collapsing for the full $4\times4$ case because there is only one possible full TX-RX subset.

We further see in Fig.~\ref{Fig4} that optimizing the ensemble of RIS configurations can substantially increase $\rho$, especially for low values of $N_\mathrm{T}=N_\mathrm{R}$. At $K_\mathrm{DUT}=100$, MP optimization increases the median $\rho$ from $4.44$ to $5.37$ in the SISO case, from $6.55$ to $7.22$ for $2\times2$, from $7.84$ to $8.36$ for $3\times3$, and from $8.62$ to $8.97$ for $4\times4$. FU optimization provides a further but smaller increase, reaching $5.66$, $7.40$, $8.52$, and $9.08$ for $1\times1$, $2\times2$, $3\times3$, and $4\times4$, respectively. Thus, MP optimization reaps most of the available improvement over random RIS selection, while FU optimization mainly gives an incremental gain.

We also checked the robustness of our multistart coordinate-exchange search. The six initial ensembles are useful because the best retained start is not always the structured one: for the MP search, the greedy initialization is retained in $48\%$ of the optimizations; for the FU search, the MP-warm initialization is retained in $62\%$ of the optimizations.
Although the median gain from choosing the best of the six starts rather than the structured start is essentially zero, the tail is non-zero, with 90th-percentile gains of $0.032$ and $0.0070$ in $\rho$ for MP and FU, respectively. The coordinate-exchange passes are close to convergence by the fourth pass: the fourth-pass median gain is zero for both MP and FU, and the 90th-percentile fourth-pass gain is zero for MP and $1.2\times10^{-4}$ for FU. These diagnostics support our choice to use six initializations each with four coordinate-exchange passes.

\subsection{Analysis of Multiplexed DUT De-Embedding}
\label{subsec_AnalysisDUTdeembedding}

\begin{table*}
    \centering
    \caption{Median and 10th-to-90th percentiles for $K_\mathrm{cal}=1024$ and $K_\mathrm{DUT}=100$. The table reports $\zeta_\mathrm{stability}$, evaluated over both OTA-characterization and DUT measurements, and the MSE of $\widehat{\mathbf S}^{\mathrm{DUT}}$ for random and MP-optimized RIS configurations.}    
    \label{Table3}
    \begin{tabular}{c c c c}
        \hline
        $N_\mathrm{T}=N_\mathrm{R}$ & $\zeta_\mathrm{stability}$ [dB] & MSE w/ random RIS configurations & MSE w/ MP-optimized RIS configurations \\
        \hline
        1 & 42.9 \textcolor{gray}{(37.6--46.4)} & $1.57{\times}10^{-3}$ \textcolor{gray}{($5.27{\times}10^{-4}$--$1.27{\times}10^{-2}$)} & $8.59{\times}10^{-4}$ \textcolor{gray}{($4.81{\times}10^{-4}$--$8.48{\times}10^{-3}$)} \\
        2 & 43.3 \textcolor{gray}{(40.9--45.5)} & $5.79{\times}10^{-4}$ \textcolor{gray}{($2.74{\times}10^{-4}$--$1.30{\times}10^{-3}$)} & $5.42{\times}10^{-4}$ \textcolor{gray}{($2.81{\times}10^{-4}$--$1.24{\times}10^{-3}$)} \\
        3 & 43.5 \textcolor{gray}{(42.0--44.6)} & $5.82{\times}10^{-4}$ \textcolor{gray}{($3.62{\times}10^{-4}$--$1.06{\times}10^{-3}$)} & $4.84{\times}10^{-4}$ \textcolor{gray}{($3.01{\times}10^{-4}$--$9.36{\times}10^{-4}$)} \\
        4 & 43.4 \textcolor{gray}{(43.4--43.4)} & $2.33{\times}10^{-4}$ \textcolor{gray}{($2.16{\times}10^{-4}$--$2.94{\times}10^{-4}$)} & $2.36{\times}10^{-4}$ \textcolor{gray}{($2.27{\times}10^{-4}$--$2.90{\times}10^{-4}$)} \\\hline  
    \end{tabular}
\end{table*}

Finally, we now evaluate our multiplexed DUT de-embedding on three reciprocal four-port DUTs. DUT~\textcircled{1} is a coupled transmission-line network, whereas DUT~\textcircled{2} and DUT~\textcircled{3} are ensembles of individual delay lines with approximately diagonal scattering matrices. In all cases, we estimate a general reciprocal four-port scattering matrix (i.e., ten independent complex-valued coefficients) and use the directly measured DUT scattering matrices only to compute the reconstruction mean-squared error (MSE), evaluated entrywise over the full DUT scattering matrix.

For a given TX-RX subset choice and DUT, we evaluate the MSE for the MP-optimized RIS ensemble and for three random RIS ensembles. The random-baseline MSE statistics pool the retained MSE values over the three random ensembles, the three DUTs, and all possible TX-RX subset choices; the MP-optimized MSE statistics pool the retained MSE values over the three DUTs and all possible TX-RX subset choices.

Before analyzing the reconstruction error, we summarize the experimental stability across both OTA-characterization and DUT measurements in Table~\ref{Table3}. While we reported the worst-checkpoint stability values across the relevant sentinel measurements during OTA characterization in Table~\ref{Table1}, the stability values reported in Table~\ref{Table3} include both the OTA-characterization and DUT measurements. The median stability values range from $42.9$ to $43.5$~dB and are therefore roughly 10~dB lower than the approximately $53$~dB stability values reported in Table~\ref{Table1}. We attribute this reduction to slow experimental drift; its impact is more apparent for Table~\ref{Table3} because that table covers a longer sequence of measurements. Nevertheless, the stability remains sufficiently high for a meaningful analysis of the reconstruction trends.

\begin{figure}
    \centering
    \includegraphics[width=\columnwidth]{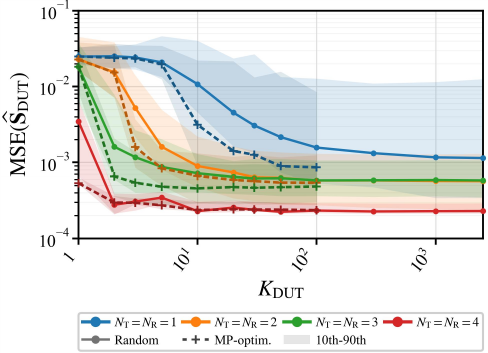}
    \caption{Multiplexed DUT-de-embedding error as a function of the number of DUT measurements $K_\mathrm{DUT}$ for $K_\mathrm{cal}=1024$. Solid lines show random RIS ensembles, and dashed lines show MP-optimized RIS ensembles selected using $\rho$. Lines show medians, and shaded regions indicate 10th-to-90th-percentile intervals across the evaluated TX-RX subset choices and DUT cases.}    
    \label{Fig5}
\end{figure}

We plot in Fig.~\ref{Fig5} the MSE of the de-embedded DUT scattering matrix as a function of the number of DUT measurements $K_\mathrm{DUT}$. For both random and MP-optimized ensembles of RIS configurations, the MSE decreases once $K_\mathrm{DUT}$ is large enough, and this transition occurs earlier for larger values of $N_\mathrm{T}=N_\mathrm{R}$. This behavior is qualitatively consistent with the necessary counting condition in~\eqref{eq_DUT_counting_condition}. Since a reciprocal four-port DUT has $d=10$ independent complex-valued coefficients, this condition requires at least $K_\mathrm{DUT}=10$, $3$, $2$, or $1$ measurement(s) for $N_\mathrm{T}=N_\mathrm{R}=1$, $2$, $3$, or $4$, respectively. These thresholds do not guarantee accurate de-embedding, but they indicate the minimal measurement set that can in principle contain enough scalar observations to identify the DUT.

The largest benefit of MP optimization occurs near these transition regions. For instance, MP optimization reduces the median MSE by $5.36$~dB for SISO at $K_\mathrm{DUT}=10$, by $5.17$~dB for $2\times2$ at $K_\mathrm{DUT}=3$, by $3.89$~dB for $3\times3$ at $K_\mathrm{DUT}=2$, and by $8.11$~dB for $4\times4$ at $K_\mathrm{DUT}=1$. Once $K_\mathrm{DUT}$ is large relative to the corresponding counting threshold, both the marginal benefit of increasing $K_\mathrm{DUT}$ and the effect of MP optimization become smaller. This saturation thus occurs earliest for $4\times4$ MIMO, where MP optimization substantially improves the median MSE only at $K_\mathrm{DUT}=1$. By contrast, SISO remains more sensitive to the choice of the RIS configurations even at larger $K_\mathrm{DUT}$. At $K_\mathrm{DUT}=100$, MP optimization reduces the median MSE for SISO from $1.57{\times}10^{-3}$ to $8.59{\times}10^{-4}$, and also reduces the upper tail from $1.27{\times}10^{-2}$ to $8.48{\times}10^{-3}$.

Table~\ref{Table3} summarizes all results at the $K_\mathrm{DUT}=100$ operating point. While the benefits of MP optimization at $K_\mathrm{DUT}=100$ are still substantial for SISO (as discussed above), the median improvements at $K_\mathrm{DUT}=100$ are more modest for $2\times2$ MIMO and $3\times3$ MIMO, from $5.79{\times}10^{-4}$ to $5.42{\times}10^{-4}$ and from $5.82{\times}10^{-4}$ to $4.84{\times}10^{-4}$, respectively. For $4\times4$ MIMO, random RIS configurations already yield an MSE of $2.33{\times}10^{-4}$, and MP optimization does not improve this value. Overall, these results show that optimizing the RIS configurations is most useful when the number of accessible antennas constitutes a bottleneck in terms of the acquired information relative to the required information, whereas sufficiently large MIMO measurements are generally already well conditioned with random RIS configurations.

\begin{figure}
    \centering
    \includegraphics[width=\columnwidth]{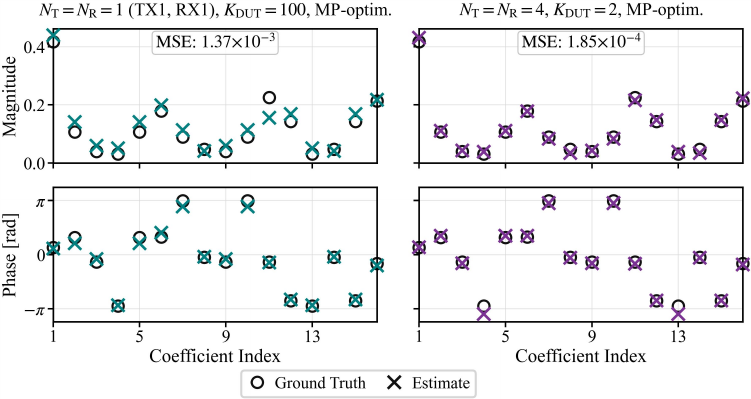}
    \caption{Representative reconstruction examples for DUT~\textcircled{1} using MP-optimized RIS configurations. Open circles indicate the ground-truth DUT coefficients, and crosses indicate the reconstructed coefficients. The SISO case uses $K_\mathrm{DUT}=100$ measurements for one representative TX-RX subset, whereas the $4\times4$ MIMO case uses only $K_\mathrm{DUT}=2$ measurements.}
    \label{Fig6}
\end{figure}

The representative reconstructions in Fig.~\ref{Fig6} illustrate the same trade-off at the level of individual DUT scattering coefficients. For DUT~\textcircled{1}, SISO de-embedding with $K_\mathrm{DUT}=100$ MP-optimized RIS configurations yields an MSE of $1.37{\times}10^{-3}$. By contrast, the $4\times4$ MIMO case achieves an MSE of $1.85{\times}10^{-4}$ with only $K_\mathrm{DUT}=2$ MP-optimized RIS configurations. This comparison highlights the central benefit of multiplexed de-embedding: RIS diversity can compensate for a small number of accessible antennas, while additional TX-RX measurements reduce the required number of RIS-programmed OTA fixture realizations.

We also checked the importance of optimizer restarts for avoiding bad local optima arising from the nonconvex DUT de-embedding problem. For the MP-optimized RIS ensembles, selecting the best of five initializations instead of the first one has negligible effect in almost all cases: the median objective-value reduction is below $10^{-5}$~dB, and even at $K_\mathrm{DUT}=100$ the 90th percentile is only $0.03$~dB. The ensembles of random RIS configurations show a similarly negligible median reduction, $5.2{\times}10^{-7}$~dB across all shown $K_\mathrm{DUT}$ values, but a larger upper tail for very small $K_\mathrm{DUT}$, where the de-embedding problem is poorly conditioned. Once $K_\mathrm{DUT}\geq20$, the 90th percentile of the reduction for the random baseline falls below $0.29$~dB, and at $K_\mathrm{DUT}=100$ it is $0.097$~dB.

\section{Conclusion}
\label{sec_conclusion}

To summarize, we introduced and experimentally validated RIS-aided wireless multiport sensing. We first characterized an RIS-programmable OTA fixture using a known TLN. Then, we measured a DUT via various realizations of the RIS-programmable OTA fixture. We performed multiplexed de-embedding of the RIS-programmable OTA fixture to estimate the DUT's scattering matrix. Our results show that RIS diversity can compensate for a limited number of accessible antennas: even in the SISO case, where each measurement provides only one scalar end-to-end channel coefficient, we recover the scattering matrix of reciprocal four-port DUTs with useful accuracy when enough realizations of the RIS-programmable OTA fixture are used. We also showed that model-based optimization of the deployed ensemble of RIS configurations is most beneficial near the transition between poorly conditioned and accurate de-embedding, with diminishing impact once the accessible antennas already provide sufficient measurement diversity.

Looking forward, an important direction is to further reduce RF-chain complexity by extending the approach to noncoherent measurements, building on the fact that related Virtual-VNA formulations can operate without coherent detection~\cite{del2024virtual,del2024virtual2p0,del2025virtual3p1}. Other potential extensions include broadband operation, adaptive RIS-configuration selection during measurements, drift compensation, and experimental implementations with distributed programmable scatterers such as RFID tags. These directions would move RIS-aided wireless multiport sensing closer to low-cost deployments in industrial, biomedical, and smart-environment sensing scenarios.

\section*{Acknowledgment}
The author acknowledges IETR's QOSC test facility (which is part of the CNRS RF-Net network).

\bibliographystyle{IEEEtran}
%\bibliography{refs}

% Generated by IEEEtran.bst, version: 1.14 (2015/08/26)
\providecommand{\noopsort}[1]{}\providecommand{\singleletter}[1]{#1}%

\end{document}